\documentclass[aps,prb,showpacs]{revtex4}

\usepackage[T1]{fontenc}
\usepackage[latin1]{inputenc}
\usepackage{graphics}

\usepackage{amsmath}


\begin{document}

\title{Spin-transfer in diffusive ferromagnet-normal metal systems with spin-flip
scattering}

\author{Alexey Kovalev,\( ^{1} \) Arne Brataas,\( ^{2} \) and Gerrit E.W. Bauer\( ^{1} \) }

\affiliation{\( ^{1} \)Department of NanoScience, Delft University of Technology,
2628 CJ Delft, The Netherlands}

\affiliation{\( ^{2} \)Harvard University, Lyman Laboratory of Physics, Cambridge, Massachusetts
02138, USA}

\begin{abstract}
The spin-transfer in biased disordered ferromagnet (F) - normal metal (N) systems
is calculated by the diffusion equation. For F1-N2-F2 and N1-F1-N2-F2-N3 spin
valves, the effect of spin-flip processes in the normal metal and ferromagnet
parts are obtained analytically. Spin-flip in the center metal N2 reduces the
spin-transfer, whereas spin-flip in the outer normal metals N1 and N3 can increase
it by effectively enhancing the spin polarization of the device. 
\end{abstract}
\date{\today{}}

\maketitle

\section{Introduction}

A spin-polarized electric current flowing through magnetic multilayers with
canted magnetizations produces torques on the magnetic moments of the ferromagnets.\cite{Sloncz:mmm96,Berger:prb96}
The effect is inverse to the giant magnetoresistance, in which a current is
affected by the relative orientation of the magnetization directions. The
spin-current-induced magnetization torque arises from an interaction between
conduction electron spins and the magnetic order parameter, transferring angular
momentum between ferromagnetic layers, hence the name \textquotedblleft spin
transfer\textquotedblright . The observed asymmetry of the switching with respect
to the direction of current flow in the magnetization switching 
in cobalt layers\cite{Myers:sc99,Katine:prl00,Grollier:apl2001,Wegrowe:jap02}
is strong evidence that the spin-transfer dominates charge current-induced Oersted magnetic
fields in mesoscopic small structures. Spin-transfer devices are promising for
applications by the ability to excite and probe the dynamics of magnetic moments
at small length scales. Reversing magnetizations with little power consumption can be utilized in current-controlled
magnetic memory elements. As a result the spin-transfer effect has
already been the subject of several theoretical studies. 
\cite{Bazaliy:prb98,Brataas:prl00,Waintal:prb00,Brataas:epjb01,Xia:prb01,Xia:prb02,Stiles:jap02,Stiles:prb02,Huertas:prb00}

\begin{figure}
\centerline{\includegraphics{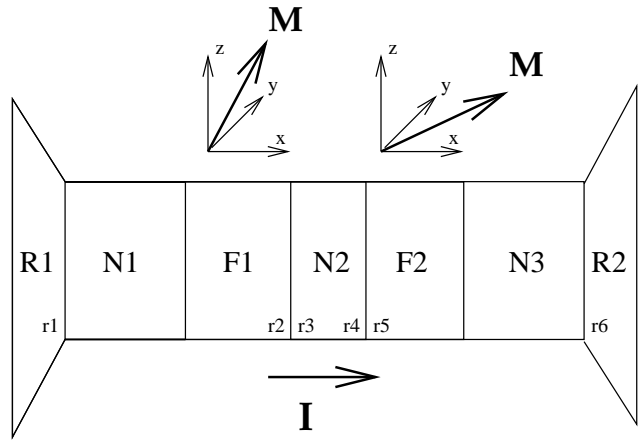}}
\caption{N1-F1-N2-F2-N3 multilayer system with non-collinear magnetizations.}

\label{fig1}
\end{figure}

The torque can be formulated by scattering theory in terms of the spin-dependence
of the reflection coefficients of the interface and the incoherence of spin-up
and spin-down states inside the ferromagnet. This leads to destructive interference
of the component of the spin-current perpendicular to the magnetization over
the ferromagnetic decoherence length, which is smaller than the mean free path
for not too weak ferromagnets.\cite{Brataas:prl00,Waintal:prb00,Brataas:epjb01,Xia:prb01,Xia:prb02,Stiles:jap02,Stiles:prb02}
In this paper we solve the spin-dependent diffusion equation for a multilayer
system consisting of two reservoirs, three normal metal layers and two ferromagnetic
layers, see Fig. \ref{fig1}, generalizing the approach of Valet and Fert
\cite{Valet:prb93} to non collinear systems.

We focus on relatively large systems in which the bulk resistance
dominates.  Interfaces play an essential role in transferring the
torque, but are assumed not significantly to increase the total
resistance. A typical interface resistance for \textit{e.g.} Co/Cu is
\( AR_{\text {surface}}\sim \mathrm{f}\Omega \mathrm{m}^{2} \).  The
corresponding typical bulk resistance for clean and dirty Co/Cu layer
varies between \( AR_{\text {bulk}}=0.01L [\mathrm{nm}] \) \(
\mathrm{f}\Omega \mathrm{m}^{2} \) and \( 0.1L [\mathrm{nm}] \) \(
\mathrm{f}\Omega \mathrm{m}^{2} \) (see for example
Ref.~\onlinecite{Bass:mmm99} and Ref.~\onlinecite{Eid:mmm01})
where \( L \) is the length of the layers expressed in \( \mathrm{nm}
\).  Our approach is therefore quantitatively valid when the layers are
thicker than \( 100 \) \( \mathrm{nm} \) for pure samples and \( 10 \)
\( \mathrm{nm} \) for alloys which is reasonable for experimental
fabrication, and furthermore reveal qualitative effects of spin-flip relaxation
process on the spin-torques for thinner layers.  Related calculations
of the torque and the magnetoresistance for submicron Co/Cu
multilayers using the Boltzmann equation have been presented in
Ref.~\onlinecite{Stiles:jap02}.  Here we obtain analytical
results based on the diffusion equation that reveal the main physical
effects of spin-flip scattering in different parts of the multilayer
on the spin transfer: Spin-flip scattering in the middle normal metal
N2 reduces the spin-transfer whereas spin-flip scattering in the outer
normal metals N1 and N3 can enhance the spin-transfer.

The paper is organized as follows: in Section II we explain the averaging mechanisms
of spin-transfer and the boundary conditions for the diffusion equation. The
latter are formulated for a N1-F1-N2-F2-N3 multilayer system and solved analytically
in the presence of spin-flip processes in the bulk layers in Sections III and IV. In
the last section we summarize our conclusion. In Appendix A, magnetoelectronic
circuit theory\cite{Brataas:prl00,Brataas:epjb01} is shown to be consistent
with the results from the diffusion equation in the absence of spin-flip scattering.

\section{Diffusive approach to multilayer systems}

Electron states with spins that are not collinear to the magnetization direction
are not eigenstates of a ferromagnet, but precess around the magnetization vector.
In three dimensions, a non-collinear spin current is composed by many states
with different Larmor frequencies which average out quickly in a ferromagnet as
a function of penetration depth. The efficient relaxation of the non-diagonal
terms in the spin-density matrix is equivalent to the suppression of spin-accumulation
non-collinear to the magnetization in the ferromagnet.\cite{Brataas:prl00,Brataas:epjb01,Stiles:jap02,Stiles:prb02}
This spin-dephasing mechanism does not exist in normal metals, in which the
spin wave functions remain coherent on the length scale of the spin-diffusion
length, which can be of the order of microns. In ballistic systems, the spin-transfer
occurs over the ferromagnetic decoherence length \( \lambda _{c}=1/\left| k_{F}^{\uparrow }-k_{F}^{\downarrow }\right|  \).
In conventional ferromagnets the exchange energy is of the same order of magnitude
as the Fermi energy and \( \lambda _{c} \) is of the order of the lattice constant.
The strongly localized regime in which the mean free path is smaller than the inverse
Fermi wave-vector, \( \ell <1/k_{F} \), is not relevant for elemental metals.
In conventional metallic ferromagnets \( \ell \gg 1/k_{F} \), and the length scale
of the spin-transfer \( \lambda _{c} \) is necessarily smaller than the mean
free path \( \ell  \) and therefore is not affected by disorder. This argument does not hold
for gradual interfaces and domain walls. The opposite
limit has been considered in Ref.~\onlinecite{Zhang:prl02},
where \( \lambda _{c}=\sqrt{2hD_{0}/J} \) (in the paper it
was designated by \( \lambda _{J} \)), or with \( \displaystyle D_{0}\sim \frac{\ell ^{2}}{\tau } \),
\( \lambda _{c}\sim \ell \sqrt{2h/J\tau } \). The limit considered
in Ref.~\onlinecite{Zhang:prl02} implies \( 2h/J\tau \gg 1 \) or \( \lambda _{c}\gg \ell  \) and
therefore does not hold for ferromagnetic conductors like Fe, Co, Ni and its
alloys.

Semiclassical methods cannot describe processes on length scales smaller than
the mean free path, thus cannot properly describe abrupt interfaces. It is possible,
however, to express boundary conditions in terms of transmission and reflection
probabilities which connect the distribution functions on both sides of an interface,
and have to be computed quantum mechanically.\cite{Schep:prb97} For transport,
these boundary conditions translate into interface resistances, which arise
from discontinuities in the electronic structure and disorder at the interface.
This phenomenon has been also extensively studied in the quasi-classical theory
of superconductivity \cite{Lambert:jp98}, where a generalized diffusion approach can be used in the bulk of the
superconductor, but proper boundary conditions must be used at the interfaces
between a superconductor and another normal or superconducting metal.

These effects
can be taken into account by first-principles band-structure calculations.\cite{Schep:prb97}
In collinear systems it is possible to circumvent the problem by replacing the
interfaces by regions of a fictitious bulk material, the resistances of which
can be fitted to experiments. This is not possible anymore when the magnetizations
are non-collinear, however, because potential steps are essential for the description
of the dephasing of the non-collinear spin current and the torque.

We wish to model the multilayer system Fig.~\ref{fig1} by the diffusion equation
and interface boundary conditions. Let \( \widehat{f}(\varepsilon ) \) be the \( 2\times 2 \)
distribution matrix at a given energy \( \varepsilon  \) and \( \widehat{I} \)
the \( 2\times 2 \) current matrix in spin-space. It is convenient to expand
these matrices into a scalar particle and a vector spin-contribution \( \widehat{f}=\widehat{1}f_{0}+\hat{\sigma }\cdot \mathbf{f}_{s} \),
\( \widehat{I}=(\widehat{1}I_{0}+\hat{\sigma }\cdot \mathbf{I}_{s})/2 \). For
normal metals \( \widehat{f}^{N}=\widehat{1}f_{0}^{N}+\hat{\sigma }\cdot \mathbf{s} \),
where \( f_{0}^{N} \) is the local charge-related chemical potential and \  the
spin distribution function has magnitude \( f_{s}^{N} \) and direction \( \mathbf{s} \).
In the ferromagnet \( \widehat{f}^{F}=\widehat{1}f_{0}^{F}+\hat{\sigma }\cdot \mathbf{m}f_{s}^{F}=\widehat{1}(f_{\uparrow }+f_{\downarrow })/2+\hat{\sigma }\cdot \mathbf{m}(f_{\uparrow }-f_{\downarrow })/2 \),
where \( f_{\uparrow } \) and \( f_{\downarrow } \) are the diagonal elements
of the distribution matrix when the spin-quantization axis is parallel to the
magnetization in the ferromagnet \( \mathbf{m} \).

The diffusion equation describes transport in both the normal metal and the
ferromagnet. We first consider a single interface and disregard spin-flip scattering.
The particle and spin currents in the normal metal with diffusion constant \( D \)
are \( j=D\partial _{x}f_{0}^{N} \) and \( \mathbf{j}^{N}_{s}=D\partial _{x}\mathbf{f}_{s}^{N} \)
respectively. The particle and spin currents are conserved: \begin{equation}
\label{current conservation1}
D\partial _{x}^{2}f_{0}^{N}=0\, ,\hspace {1cm}D\frac{\partial ^{2}}{\partial x^{2}}\mathbf{f}_{s}^{N}=0\, .
\end{equation}
 In the ferromagnet the particle and spin currents are \( j=D_{\uparrow }\partial _{x}f_{\uparrow }+D_{\downarrow }\partial _{x}f_{\downarrow } \)
and \( \mathbf{j}^{F}_{s}=\mathbf{m}\partial _{x}(D_{\uparrow }f_{\uparrow }-D_{\downarrow }f_{\downarrow }) \),
where \( D_{\uparrow } \) and \( D_{\downarrow } \) are the diffusion constants
for spin-up and spin-down electrons. Current conservation of the spin components
parallel and antiparallel to the magnetization direction in the ferromagnet
read: \begin{equation}
\label{current conservation2}
D_{\uparrow }\partial _{x}^{2}f_{\uparrow }=0\, ,\hspace {1cm}D_{\downarrow }\partial _{x}^{2}f_{\downarrow }=0\, .
\end{equation}
 Eqs. (\ref{current conservation1},\ref{current conservation2}) are applicable
only inside the bulk layers. The boundary conditions at the interface arise
from the continuity of the particle and spin distribution functions on the normal
and the ferromagnetic metal sides:\cite{Brataas:prl00,Brataas:epjb01}\begin{equation}
\label{continuity}
f_{s}^{N}|_{\text {N-surface}}=(f_{\uparrow }+f_{\downarrow })/2|_{\text {F-surface}}\,,
\end{equation}
\begin{equation}
\label{continuity1}
\mathbf{f}_{s}^{N}|_{\text {N-surface}}=\mathbf{m}(f_{\uparrow }-f_{\downarrow })/2|_{\text {F-surface}}\, .
\end{equation}
 Furthermore, particle current is conserved:\cite{Brataas:prl00,Brataas:epjb01}\begin{equation}
\label{current boundary}
\lbrack D\partial _{x}f_{0}^{N}]|_{\text {N-surface}}=\partial _{x}(D_{\uparrow }f_{\uparrow }+D_{\downarrow }f_{\downarrow })|_{\text {F-surface}}\, .
\end{equation}
 We have discussed above why the non-collinear component of the spin-accumulation
decays on a length scale of the order of the lattice spacing. This leads to
the third boundary condition at the F-N interface, namely that the spin-current
is conserved only for the spin-component parallel to the magnetization direction:\cite{Brataas:prl00,Brataas:epjb01}\begin{equation}
\label{spincurrent boundary}
\lbrack D\partial _{x}\mathbf{f}_{s}]|_{\text {N-surface}}={\mathbf{m}}\partial _{x}(D_{\uparrow }f_{\uparrow }-D_{\downarrow }f_{\downarrow })|_{\text {F-surface}}\, .
\end{equation}
 Solving these equations, we recover Eq. (\ref{spintorque}) with the mixing
conductance as found by the magnetoelectric circuit theory (Appendix A).\cite{Brataas:prl00,Brataas:epjb01} The magneto-electronic circuit theory is thus equivalent to the diffusion approach when the system size is larger than the mean free path. However the  magneto-electronic circuit theory is a more general approach that can also be used for circuits or parts of circuits that are smaller than the mean free path.
Note that the boundary conditions above do not contain explicit reference to interface
conductance parameters and are therefore valid only for bulk resistances which
are sufficiently larger than the interface resistances. The gain by using the diffusion equation is,
that we can now easily derive simple analytical results, also in the presence of spin-flip
relaxation. In the normal as well as ferromagnetic
metals spin-flip scattering leads to:\begin{equation}
\label{spin-flip}
\partial _{x}j_{0}=0\, ,\hspace {1cm}\frac{\partial }{\partial x}\mathbf{j}_{s}=\mathbf{f}_{s}/\tau _{sf}\, ,
\end{equation}
 where the spin-flip relaxation time \( \tau _{sf} \) is a material dependent
parameter.

\section{Results for systems without spin-flip}

Let us now apply this method to the spin-transfer in a N1-F1-N2-F2-N3 system
Fig.~\ref{fig1} to obtain explicit results for the figure of merit, \textit{viz}.
the ratio of the spin-torque to the charge current through (or voltage bias across) the system. The layers
are characterized by the lengths \( L_{N1} \), \( L_{F1} \), \( L_{N2} \),
\( L_{F2} \), \( L_{N3} \) and by diffusion constants \( D_{N1} \), \( D_{F1,\uparrow (\downarrow )} \),
\( D_{N2} \), \( D_{F2,\uparrow (\downarrow )} \), \( D_{N3} \) for each
normal and ferromagnetic metal layer, respectively. The resistances of the system
are \( R_{N1} \), \( R_{F1,\uparrow } \), \( R_{F1,\downarrow } \), \( R_{N2} \),
\( R_{F2,\uparrow } \), \( R_{F2,\downarrow } \) and \( R_{N3} \) with, for
example, \( R_{N1}=L_{N1}/(A_{N1}D_{N1}) \) and \( R_{F1,\uparrow }=L_{F1}/(A_{F1}D_{F1,\uparrow }) \)
(\( L \) and \( A \) are length and cross section of a layer respectively).
Let us initially disregard spin-flip scattering.

The continuity of the spin-current at the interface N1-F1 can easily be shown from Eqs. (\ref{current conservation1},\ref{current conservation2},\ref{continuity},\ref{current boundary}).
As a result the two layers N1-F1 behave effectively like a single ferromagnetic layer with renormalized
resistance: \begin{subequations}
\label{renorm}
\begin{eqnarray}
\widetilde{R}_{F1,\uparrow } &=&R_{F1,\uparrow }+2R_{N1}\,, \\
\widetilde{R}_{F1,\downarrow } &=&R_{F1,\downarrow }+2R_{N1}\,.
\end{eqnarray} \end{subequations}The same is true for the interface F2-N3. As
a result it is sufficient to treat only the F1-N-F2 system. In general, there
are spin-current discontinuities at the interfaces F1-N and N-F2 which, due to
momentum conservation, lead to torques acting on the magnetic moments in
the ferromagnetic layers. Taking into account all diffusion equations (\ref{current conservation1},\ref{current conservation2})
and boundary conditions (\ref{continuity},\ref{continuity1},\ref{current boundary},\ref{spincurrent boundary}),
and also introducing the parameters \( R=R_{N2} \), \( R_{i\pm }=(\widetilde{R}_{Fi,\uparrow }\pm \widetilde{R}_{Fi,\downarrow })/4 \),
where \( i=1,2 \), the torques can be written as \\
 \begin{subequations}
\label{torquea}
\begin{eqnarray}
\tau _{1} =I_{0}R_{2-}\frac{\left( R+R_{1+}-\alpha
R_{1-}R_{2+}/R_{2-}\right) }{(R+R_{2+})(R+R_{1+})-\alpha ^{2}R_{1+}R_{2+}}(\alpha \mathbf{m}_{1}-\mathbf{m}_{2})\,, \\
\tau _{2} =I_{0}R_{1-}\frac{R+R_{2+}-\alpha 
R_{2-}R_{1+}/R_{1-}}{(R+R_{2+})(R+R_{1+})-\alpha ^{2}R_{1+}R_{2+}}(\mathbf{m}_{1}-\alpha \mathbf{m}_{2})\,,
\end{eqnarray}\end{subequations} where \( \tau _{1} \) and \( \tau _{2} \)
are torques acting on the magnetizations of the first and second ferromagnet
respectively, \( \alpha = \)\textbf{\( (\mathbf{m}_{1}\cdot \mathbf{m}_{2}) \)}\( =\cos \theta  \),
\( \theta  \) being the angle between the magnetizations. The resistance can
also be calculated: \begin{eqnarray}
\Re (\theta )=R+R_{1+}+R_{2+}- \label{mresistance1} 
\frac{R^{2}_{1-}+2\alpha R_{1-}R_{2-}+R^{2}_{2-}+(1-\alpha ^{2})(R^{2}_{1-}R_{2+}+R^{2}_{2-}R_{1+})/R}{R+R_{1+}+R_{2+}+R_{1+}R_{2+}(1-\alpha ^{2})/R} \,  .
\end{eqnarray}
 It is worthwhile to rewrite the above formula (\ref{torquea}) using the effective
polarization \( P=R_{-}/R_{+} \) (which is the polarization of a current flowing
through F or N-F layers connected to reservoirs) and the ferromagnet charge
current resistance \( R_{i}=R_{i+} \). The (absolute values of the) torques
are then: \begin{subequations}
\label{torque}
\begin{eqnarray}
\left\vert \tau _{1}\right\vert  &=&\frac{\left\vert 1+R/R_{1}-\alpha
P_{1}/P_{2}\right\vert \left[ (1-\alpha ^{2})I_{0}P_{2}\right] 
}{(1+R/R_{2})(1+R/R_{1})-\alpha ^{2}}\,, \\
\left\vert \tau _{2}\right\vert  &=&\frac{\left\vert 1+R/R_{2}-\alpha
P_{2}/P_{1}\right\vert \left[ (1-\alpha ^{2})I_{0}P_{1}\right] 
}{(1+R/R_{1})(1+R/R_{2})-\alpha ^{2}}\,.
\end{eqnarray}\end{subequations} As one can see from Eqs. (\ref{torquea}) and
(\ref{torque}) there is an asymmetry with respect to current inversion. For
example, if only one polarization can rotate (one ferromagnet is much wider
than the other or exchange-biased), domains in the two magnetic layers can be aligned antiparallel
by currents flowing in one direction, and reoriented parallel by reversing the
current flow. This happens because only one state (parallel or
antiparallel) is at equilibrium for a fixed direction of the current. If the
currents are large enough (depending on other sources of torques such as external
fields, magneto crystalline anisotropy and damping) the magnetization will flip,
which can be monitored by the change in total resistance of \begin{equation}
\label{difference}
\frac{R(\uparrow \downarrow )-R(\uparrow \uparrow )}{R(\uparrow \downarrow )}=\frac{4R_{1-}R_{2-}}{R^{2}+(R_{1+}+R_{2+})^{2}-(R_{1-}+R_{2-})^{2}}\, .
\end{equation}
 In the case of unit polarization and \( R\approx 0 \) the relative
resistance change (\ref{difference}) can be 100 percent. This asymmetry was
predicted by the spin-transfer theory \cite{Sloncz:mmm96} and was observed
experimentally. \cite{Myers:sc99,Katine:prl00,Grollier:apl2001} Note, however, that in these experiments
the mean free path is comparable to the size of the systems, and the present
theory cannot be directly applied.

\begin{figure}
\centerline{\includegraphics{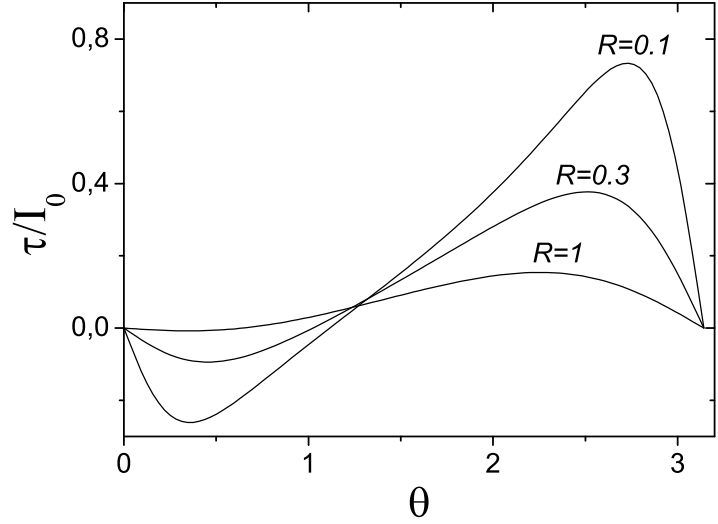}}

\caption{Torque acting on the first ferromagnet as a function of the relative angle between
the two magnetizations for different normal metal resistances (the resistances
are expressed in units \protect\protect\protect\protect\( R_{1}=R_{2}\protect \protect \protect \protect \),
\protect\( P_{1}=0.5\protect \),
\protect\( P_{2}=0.2\protect \)). }

\label{fig2}\par{} \vspace{0.3cm}
\end{figure}

From Eq. (\ref{torque}) follows that the torques are equal to zero for parallel
and antiparallel alignments. When the numerator of Eq. (\ref{torque}) \( 1+R/R_{1(2)}-\alpha P_{1(2)}/P_{2(1)} \)
never vanishes, the torque increases with \( \theta  \) from zero to a maximal
value which corresponds to an angle larger than \( \pi /2 \) and vanishes again
when configurations become antiparallel. When the nominator of Eq. (\ref{torquea})
does vanish for some angle \( \theta _{0} \), the absolute value of torque has
a local maximum before \( \theta _{0} \) (see Fig. \ref{fig2}). In principle, it is possible to have
an equilibrium magnetization
angle \( \theta =\theta _{0} \) for one current direction while equilibrium
magnetization angle \( \theta =0 \) or \( \pi  \) for the opposite current
direction (this can lead to asymmetry for the transition from the anti-aligned state to the aligned state in 
comparison with the transition from aligned to anti-aligned observed experimentally \cite{Wegrowe:jap02}).

We propose a setup in which only one magnetization can rotate (usually it is
achieved by taking one ferromagnetic layer much wider than the other or by exchange biasing). If one
ferromagnetic layer (for example the first one) has a resistance \( R_{1}\ll R \)
and the other \( R_{2}>R \), the torque \( \tau _{2} \) vanishes whereas the
other torque can be simplified to: \begin{equation}
\tau _{1}=(1-\alpha ^{2})I_{0}P_{2}
\end{equation}
 The maximal torque in this setup occurs when the magnetizations of the ferromagnet
F1 and the ferromagnet F2 are perpendicular.

In general, the spin-torque is maximal when the resistance \( R \)
of the normal metal vanishes, as could have been expected since this also gives
the maximum magnetoresistance effect. In Eqs. (\ref{torquea}) and (\ref{torque})
the size of the magnets does not play a dominant role for small normal metal
resistances. In this case the torques depend mainly on the polarizations.

\section{Results for systems with spin-flip}

So far, we have disregarded spin-flip scattering, which can be included readily,
though. Here the system N1-F1-N2-F2-N3 is analyzed and spin-flip in each normal
metal part is considered separately. Introducing spin-flip in N1 and N3 leads
to the simple result:
Eq. (\ref{renorm}) without spin-flip remains valid, but with modified spin resistances:\begin{equation}
\label{renormalization}
R_{N1(N3)}^{sf}=R_{N1(N3)}\frac{\tanh (L_{N1(N3)}/\ell _{sd})}{(L_{N1(N3)}/\ell _{sd})}\, ,
\end{equation} where  \( \ell _{sd} \) is the normal metal spin-flip diffusion length. 
 When \( L\gg \ell _{sd} \), the resistance is governed by the spin-flip diffusion
length \( \ell _{sd} \), which means that only part of the metal takes part
in the spin transfer whereas the rest plays the role of the reservoir. This
reduction of the active thickness of the device can lead to an effective polarization
increase by decreasing the effect of \( R_{N}  \) in Eq.(\ref{renorm}). Spin-flips in the middle normal
metal have a larger impact. The torques in
the presence of spin-flips in N2 read: \begin{subequations}
\label{torqsf}
\begin{eqnarray}
\left\vert \tau _{1}\right\vert  &=&(1-\alpha ^{2})I_{0}P_{2}\frac{\beta
+R^{sf}/R_{1}-\alpha P_{1}/P_{2}}{(\beta +R^{sf}/R_{2})(\beta
+R^{sf}/R_{1})-\alpha ^{2}}\,, \\
\left\vert \tau _{2}\right\vert  &=&(1-\alpha ^{2})I_{0}P_{1}\frac{\beta
+R^{sf}/R_{2}-\alpha P_{2}/P_{1}}{(\beta +R^{sf}/R_{1})(\beta
+R^{sf}/R_{2})-\alpha ^{2}}\, ,
\end{eqnarray}\end{subequations}where \( \beta =\cosh (L/\ell _{sd}) \) and
\( P_{1(2)} \), \( R_{1(2)} \) are given by (\ref{renorm}) and (\ref{renormalization}).
\( R^{sf} \) is a new effective normal metal resistance: \begin{equation}
\label{renormalization1}
R^{sf}=R\frac{\sinh (L/\ell _{sd})}{L/\ell _{sd}}\, .
\end{equation}
 For \( L\geq \ell _{sd} \) the torque is significantly reduced by spin-flips,
becoming exponentially small for longer samples.

Let us now consider spin-flips in the ferromagnet. The treatment of the N1-F1-N2-F2-N3
system is cumbersome, so let us concentrate on the simple case of a F-N-F system.
In that case formulas remain unchanged, provided \( R_{+} \) and \( R_{-} \)
are renormalized as: \begin{subequations}
\label{spin-flipinferromagnet}
\begin{eqnarray}
R_{1(2)-}^{sf} &=&R_{1(2)-}\frac{\tanh (L_{F1(F2)}/\ell 
_{sd}^{F})}{L_{F1(F2)}/\ell _{sd}^{F}}\,, \\
R_{1(2)+}^{sf} &=&R_{1(2)+}\frac{\tanh (L_{F1(F2)}/\ell 
_{sd}^{F})}{L_{F1(F2)}/\ell _{sd}^{F}}\,.
\end{eqnarray}\end{subequations}where  \( \ell ^{F}_{sd} \) is the ferromagnet spin-flip diffusion length. 
These resistances should be used in Eqs. (\ref{torquea})
for the torques in F-N-F systems. If spin-flip in the normal metal exists, then
the formulas (\ref{torqsf}) should be used. Eqs. (\ref{spin-flipinferromagnet})
imply that there is no polarization change (as defined below Eqs. (\ref{torquea}))
and only the ferromagnet resistances \( R_{1(2)} \) are affected. For \( L\gg \ell ^{F}_{sd} \)
the bulk of the ferromagnet behaves like a reservoir (just like for the normal
metal in the same limit) and only a slice with thickness \( \ell ^{F}_{sd} \) is
active. In general, spin-flip in the ferromagnet leads to reduced torques as
\( R_{1(2)} \) becomes smaller.

The effect may be quite small as long as the
resistance of the ferromagnet is sufficiently larger than that of the normal
metal (this can also be seen from Eqs. (\ref{renorm}) and (\ref{torque})),
so that the polarization of the current is maintained.

\begin{figure}
\centerline{\includegraphics{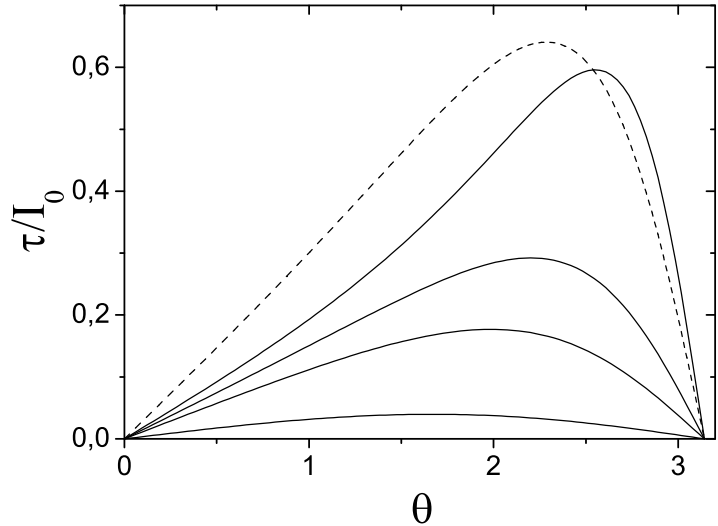}}
\caption{Torque on each ferromagnet as a function of the relative angle between the
two magnetizations for different spin-flip diffusion lengths in the normal metal
(the resistance \protect\protect\protect\protect\( R=0.4\protect \protect \protect \protect \)
is expressed in units \protect\( R_{1}=R_{2}\protect \),
\protect\( P_{1}=P_{2}=0.4\protect \protect \protect \protect \),
\protect\( L/\ell _{sd}=0\protect \protect \protect \protect \),
\protect\( 1\protect \),
\protect\( 1.5\protect \),
\protect\( 3\protect \)
the lower plot corresponds to the higher ratio). By dashed line we plotted Slonczewski's
result \cite{Sloncz:mmm96} for the same polarization. }

\label{fig3}\par{} \vspace{0.3cm}
\end{figure}

Finally we would like to discuss magnetoresistance and torque for the symmetric
\( R_{1+}=R_{2+} \) and \( R_{1-}=R_{2-} \). For the angular
magnetoresistance we extract from (\ref{mresistance1}) the formula observed
by Pratt and also shown to be universal for any disordered F-N-F perpendicular
spin valves in Ref.~\onlinecite{Gerrit:cm0205453}: \begin{equation}
\label{Pratt}
\frac{R(\theta )-R(0)}{R(\pi )-R(0)}=\frac{1-\cos \theta }{\chi (1+\cos \theta )+2}\, ,
\end{equation}
 with one parameter \( \chi  \) that is given by circuit theory\begin{equation}
\chi =\frac{1}{1-p^{2}}\frac{\left| \eta \right| ^{2}}{Re\eta }-1\, ,
\end{equation}
 in terms of the normalized mixing conductance \( \eta =2g_{\uparrow \downarrow }/g \),
the polarization \( p=(g_{\uparrow }-g_{\downarrow })/g \), and the average
conductance \( g=g_{\uparrow }+g_{\downarrow } \). As we do not take into account interface resistances, in our case the
parameters can be expressed only via bulk resistances: \( g=1/(R_{\uparrow }+R)+1/(R_{\downarrow }+R) \),
\( \eta =2/(Rg) \), \( p=2R_{-}/(2R_{+}+R) \). From Eqs. (\ref{torque}) and
(\ref{mresistance1}) the analytical expressions of the spin torque  on either
ferromagnet for current and voltage biased systems read: \begin{equation}
\label{torqueI}
\left| \tau \right| =\frac{p(\chi +1)\sin \theta }{\chi (\cos \theta +1)+2}I_{0}\, ,
\end{equation}
\begin{equation}
\label{torqueU}
\left| \tau \right| =\frac{pg}{2}\frac{\eta \sin \theta }{(\eta -1)\cos \theta +1+\eta }\frac{\mu _{l}-\mu _{r}}{2\pi }\, ,
\end{equation}
 where \( \mu _{l(r)} \) is the chemical potential in the left (right) ferromagnet.
In the presence of spin-flip for the angular magnetoresistance we can write
(restricting ourself to F-N-F case again): \begin{equation}
\label{mresistance2}
\frac{R(\theta )-R(0)}{R(\pi )-R(0)}=\frac{(1+\chi (\beta -1)/2)(1-\cos \theta )}{\chi (\beta +\cos \theta )+2}\, ,
\end{equation}
 where all parameters should be calculated according to (\ref{renormalization1})
and (\ref{spin-flipinferromagnet}). The dependences of the torque on angle
now read:\begin{equation}
\label{torque2}
\left| \tau \right| =\frac{p(\chi +1)\sin \theta }{(\chi (\cos \theta +\beta )+2)}I_{0}\, ,
\end{equation}

\begin{equation}
\label{torqueU1}
\left| \tau \right| =\frac{pg}{2}\frac{\eta \sin \theta }{A_{1}\cos \theta +A_{2}}\frac{\mu _{l}-\mu _{r}}{2\pi }\, ,
\end{equation}
 where we introduced two parameters: \begin{eqnarray}
A_{1} = -\frac{p^{2}}{1+\displaystyle\frac{\chi }{2}\frac{\beta -1}{\chi +1}}(1+\frac{\chi \beta +\chi }{2})+
\chi (1+\frac{(\kappa _{1}+\chi \kappa _{2})}{\chi +1}-\kappa _{2}p)\: ,
\end{eqnarray}

\begin{eqnarray}
A_{2}=\frac{p^{2}}{1+\displaystyle\frac{\chi }{2}\frac{\beta -1}{\chi +1}}(1-\frac{\chi \beta +2}{2}(\beta +1))+
(1+\frac{(\kappa _{1}+\chi \kappa _{2})}{\chi +1}-\kappa _{2}p)(\chi \beta +2)\: ,
\end{eqnarray}
and \( \kappa _{1}=\displaystyle\frac{L_{N}/\ell _{N,sd}}{\sinh (L_{N}/\ell _{N,sd})}-1 \),
\( \kappa _{2}=\displaystyle\frac{L_{F}/\ell _{F,sd}}{\tanh (L_{F}/\ell _{F,sd})}-1 \).

An interesting result can be drawn from Eqs. (\ref{mresistance2})
and (\ref{torque2}) by comparison with the Eqs. (\ref{Pratt}) and (\ref{torqueI}).
In order to fit the torque and the magnetoresistance in the presence of spin-flip
we need an additional parameter \( \beta  \) (\( \beta  \) was defined in Eqs. (\ref{torqsf})), 
which depends only on the spin-flip
diffusion length in the normal metal spacer. The general form of Eqs. (\ref{mresistance2})
and (\ref{torque2}) with only two important parameters seems to be valid even
in the presence of interfaces, but this has to be confirmed by future studies.
Eq. (\ref{torqueU1}) is cumbersome depending explicitely on the diffusion length in the ferromagnets.
In Fig. \ref{fig3} we have plotted result of Eq. (\ref{torque2}) for different
spin-diffusion lengths in the normal metal. The smaller diffusion
length corresponds to smaller torques. The curves only qualitatively resemble Slonczewski's
result for ballistic systems, but it should be pointed out that for the case
\( p=1 \), \( \eta =2 \) both approaches result in the same formula.

\section{Conclusion}

We investigated transport in multilayer systems in the diffusive limit with
arbitrary magnetizations in the ferromagnetic layers. The boundary conditions
for diffusion equations including spin transfer were discussed and analytic
expression for the magnetization torques and the angular magnetoresistance were
obtained. The torque can be engineered, not so much via the geometry of the
samples (like layer thicknesses) but rather via the materials, the ferromagnetic
polarization being an important parameter. The asymmetry with respect to the
current flow direction has been addressed and the resistance change under magnetization
reversal was calculated for different current directions. The effect of spin-flip
in the normal metal and ferromagnet was studied analytically. Spin-flip in the
center normal metal suppresses the spin-transfer, whereas spin-flip in the outer
normal metals can effectively increase the polarization and spin-transfer. The
spin-flip processes in the ferromagnet also diminish the spin-transfer, but
not as drastically as long as the resistance of the ferromagnet is larger than
the normal metal resistance. Finally we show in Appendix A that the diffusive
approach with carefully chosen boundary conditions leads to results which
coincide with those from circuit theory.

We are grateful to Yuli Nazarov, Yaroslav Tserkovnyak and Daniel Huertas-Hernando for stimulating
discussions. This work was supported in part by the NEDO International Joint
Research Grant Program \textquotedblleft
Nano-magnetoelectronics\textquotedblright, NSF grant DMR 99-81283 and DARPA
award No. MDA 972-01-1-0024.

\appendix

\section{Circuit theory approach to the diffusive systems}

Here we show that the diffusion approach is equivalent the circuit theory 
and that the mixing conductance is a valid concept also in
systems which are dominated by bulk transport.\cite{Tserkovnyak:prb01} We consider
a F1-N2-F2 system Fig. \ref{fig1} (N1 and N3 can also be included) connected
to two reservoirs R1 and R2 with negligible interface resistances. Note that
this does not mean that the interface is neglected, because it plays an essential
role in the boundary conditions as mentioned in the main text. Since the system
is diffusive, a thin slice of a ferromagnet or a normal metal can be considered
as a node. The mixing conductance can be written in terms of the reflection
and transmission coefficients and incorporates any kind of details of the contacts,
\textit{e.g}. tunnel, diffusive, and ballistic contacts. We are free to define
interface resistors via the location of the nodes. Here it is chosen such that
the interface width is larger than the ferromagnetic decoherence length but
smaller than the mean free path. We introduce six nodes: r1 is in R1 just before
the interface, r2 and r3 are before and after the F1-N2 interface, r4 and r5
are before and after the N2-F3 interface, r6 is in R2 just after the interface
Fig. \ref{fig1}.

Let us first find the charge and the spin current in F1 (F2) at the interface
where the spin-transfer takes place. The currents read: \begin{subequations}
\label{spinchargecurrents}
\begin{eqnarray}
I_{0} &=&(G^{\downarrow }+G^{\uparrow })(f_{0}^{F}-f_{0}^{N})+(G^{\downarrow
}-G^{\uparrow })f_{s}^{F}\,, \\
\mathbf{I}_{s} &=&\mathbf{m}\left[ (G^{\downarrow }-G^{\uparrow
})(f_{0}^{F}-f_{0}^{N})+(G^{\downarrow }+G^{\uparrow })f_{s}^{F}\right] \,,
\end{eqnarray}\end{subequations}where \( f_{0}^{N} \) is the particle distribution
function in r1 (r6) and \( f_{0}^{F} \) is the spin distribution function in
r2 (r5). The distribution function at r2 and r3 (r4 and r5) is identical due
to the continuity boundary condition. The spin-current between the ferromagnet
reservoir r2 (r5) and the normal metal reservoir r4 (r3) driven by the non-equilibrium
distributions can be found by using the circuit theory: \begin{eqnarray}
\mathbf{I}_{1(2)}=\mathbf{m}2G_{N}(f_{s}^{F}-\mathbf{s}\cdot \mathbf{m}f_{s}^{N})\label{spin-current} 
 -2\mbox {Re}G^{\uparrow \downarrow }f_{s}^{N}(\mathbf{s}-(\mathbf{s}\cdot \mathbf{m})\mathbf{m})
 +(\mathbf{s}\times \mathbf{m})2\mbox {Im}G^{\uparrow \downarrow }f_{s}^{N}\, ,
\end{eqnarray}
 where the spin accumulation in the normal metal reservoir r4 (r3) is given
by the unit vector \( \mathbf{s} \) and the spin distribution function \( f_{s}^{N} \).
Use was made of \( G^{\downarrow }=G^{\uparrow }=G_{N} \) because r2 (r5) is
close to the interface. The component of the current perpendicular to the magnetization
\( \mathbf{m} \) is transferred to the magnetization at the interface whereas
the parallel component is conserved. The torque acting on the magnetization
in F1 (F2) therefore becomes: \begin{equation}
\label{spintorque}
\tau _{1(2)}=-2\mbox {Re}G^{\uparrow \downarrow }f_{s}^{N}(\mathbf{s}^{N}-(\mathbf{s}^{N}\cdot \mathbf{m})\mathbf{m})+(\mathbf{s}^{N}\times \mathbf{m})2\mbox {Im}G^{\uparrow \downarrow }f_{s}^{N}\, .
\end{equation}
 The mixing conductance is related to the reflection coefficients of an electron
from the normal metal to the ferromagnet: \begin{equation}
\label{g}
G_{\uparrow \downarrow }=\sum _{nm}[\delta _{nm}-(r_{nm}^{\uparrow })^{\ast }r_{nm}^{\downarrow }]\, .
\end{equation}
 Let us now evaluate the mixing conductance for a disordered system. We assume
that the junction consists of two connected parts. The normal metal section
is described by a single scattering matrix for both spin-\( \uparrow  \) and
spin-\( \downarrow  \) electrons. The ferromagnetic section requires two independent
scattering matrices, one for spin-\( \uparrow  \) and one for spin-\( \downarrow  \)
electrons. Scattering at the F-N boundary is disregarded here since it is assumed
that the total resistance is dominated by the diffuse normal metal and ferromagnetic
metal parts of the junction. The total reflection matrix \( r^{\alpha } \)
for spin-\( \alpha  \) electrons can then be found by concatenating the normal
metal and ferromagnetic parts as: \begin{equation}
\label{r}
r^{\alpha }=r_{N}+t_{N}^{\prime }r_{F}^{\alpha }\sum _{n=0}^{\infty }(r_{N}^{\prime }r_{F}^{\alpha })^{n}t_{N}\equiv r_{N}+\chi ^{\alpha }\, .
\end{equation}
 By inserting (\ref{r}) into the definition for the mixing conductance we find
that the mixing conductance can be expressed as \( G_{\uparrow \downarrow }=G_{N}+\delta G_{\uparrow \downarrow } \),
where \begin{equation}
\label{mixcorr}
\delta G_{\uparrow \downarrow }=\sum _{nm}[(r_{N})_{nm}^{\ast }\chi _{nm}^{\downarrow }]+(\chi _{nm}^{\uparrow }(r_{N})_{nm})^{\ast }+(\chi _{nm}^{\uparrow })^{\ast }\chi _{nm}^{\downarrow }]\, .
\end{equation}
 Eq. (\ref{mixcorr}) depends on the phase difference between the scattering
paths of spin-up and spin-down electrons. It is assumed that there are no correlations
between the scattering matrices of the spin-\( \uparrow  \) and spin-\( \downarrow  \)
electrons in the ferromagnetic part, which is consistent with the a small coherence
length. Consequently, in a diffusive systems \( \delta G_{\uparrow \downarrow }=0 \).
However, the up- and down-spin parts of the total scattering matrix of the combined
normal metal and ferromagnetic system \textit{are} correlated since both spin
directions see the same scattering centers in the normal metal part. This leads
to the conclusion that for diffusive hybrid system: \begin{equation}
\label{diff conductance}
G_{\uparrow \downarrow }^{D}=G_{N}\, .
\end{equation}
 From Eqs. (\ref{spinchargecurrents},\ref{spin-current},\ref{spintorque})
and taking into account Eq. (\ref{diff conductance}), and noting that \( 2G_{N}=1/R \)
and \( G_{\downarrow }=1/R_{\downarrow },\, \, G_{\uparrow }=1/R_{\uparrow } \)
one can easily find the Eq. (\ref{torque}).

\bibliography{central}

\end{document}